\documentclass[12pt]{article}
\pdfoutput=1
\usepackage{amsmath,amssymb,amsfonts,amsthm,bm,bbm,cancel,wasysym}
\usepackage{epsfig,graphics,graphicx,epstopdf,caption,subcaption}
\usepackage[numbers,sort&compress]{natbib}
\graphicspath{{Charts/}}
\usepackage{array,booktabs,colortbl,colordvi,multirow}
\usepackage{colordvi,color,xcolor}
\usepackage{hyperref}
\usepackage{rotating}
\usepackage{comment}
\usepackage{feynmf}

\begin{document}

\begin{center}

{\Large {\bf Complex Field Inflation}}\\

\vspace*{0.75cm}

{Ruopeng Zhang and Sibo Zheng}

\vspace{0.5cm}
{Department of Physics, Chongqing University, Chongqing 401331, China}

\end{center}
\vspace{.5cm}

\begin{abstract}
 \noindent 
We report first study of complex field inflation.
Although understood as a specific two-field inflation,
a complex field inflation is able to make more robust model predictions on primordial curvature perturbation. 
Explicitly we discuss the model realizations of complex chaotic and exponential inflation in various large-field contexts.
Both complex field models contain a single complex scalar together with only two free parameters.
Using numerical handles aimed to calculate primordial curvature perturbation from multifield inflation, 
we show that both models are compatible with current Planck data, 
and the individual surviving parameter space can be substantially or fully probed by future CMB-S4 experiments.

\end{abstract}

\renewcommand{\thefootnote}{\arabic{footnote}}
\setcounter{footnote}{0}
\thispagestyle{empty}
\vfill
\newpage
\setcounter{page}{1}

\section{Introduction}
Observations from cosmic microwave background (CMB) experiments \cite{Hinshaw:2012aka, Planck:2018jri, Planck:2018vyg, Planck:2019kim} have placed strong constraints on amplitude, spectrum index and non-Gaussianty of primordial curvature perturbation originated from inflation \cite{Guth:1980zm, Linde:1981mu, Albrecht:1982wi} - a paradigm to resolve both horizon and flatness problem in early Universe.
These measurements are expected to be improved in higher precisions by 
future experiments such as cosmic microwave background (CMB-S4) \cite{Abazajian:2016yjj} as well as large scale structure (LSS) \cite{Alvarez:2014vva} 
which can access to energy frontiers far beyond reaches of foreseeable ground-based colliders.

So far, the measurements on the primordial curvature perturbation have provided us remarkable insights on single-field inflation.
In this context, a critical fact - the primordial large-scale curvature  perturbation is frozen \cite{Bardeen:1980kt,Lyth:2004gb} between horizon crossing and re-entering either in a radiation- or matter-dominated period - allows us to directly compare a fundamental single-field model to the CMB data. For example, the latest Planck 2018 data strongly disfavors sing-field chaotic inflation with index larger than two,
and a combined Planck 2018 plus BK15 data has already excluded single-field natural inflation \cite{Planck:2018jri}, among others.

Nevertheless, the established correspondence in the single-field case is no longer valid in multifield case whether the multifield inflation is slow-roll or not.
Because the presence of new isocurvature (entropy) \cite{Sasaki:1995aw, GarciaBellido:1995qq} perturbation in multifield inflation 
leads to a non-conversation of the curvature (adiabatic) perturbation during inflation,
as long as classical trajectory in field space is not a straight line. 
As a result, a varying curvature perturbation complexes the direct comparison between a multifield inflation and the CMB data.
The situation is further worsen by a larger number of model parameters in multifield than single-field case. Because it makes theoretical predictions of the primordial curvature perturbations in the former case less robust.
In this study we consider the question of how powerful predictions on the primordial curvature perturbation can be regained in the multifield context.

As an attempt to this question,
in this work we advocate complex field inflation from a perspective different from earlier attempts \cite{Easther:2013rva, Dias:2016slx, Dias:2017gva, Bjorkmo:2017nzd}.
A complex field inflation naturally takes place in various contexts of new physics such as supersymmetry, 
supergravity and superstring.
Although understood as a specific two-field version,
a complex field inflation is able to make more robust predictions on the primordial curvature perturbation,
as the number of free parameters therein can be obviously reduced in virtue of rational symmetry.
The rest of this work is organized as follows.
In Sec.2,  we briefly review the primordial curvature perturbation in complex field inflation.
Then, in Sec.3 we consider model realizations of complex chaotic and exponential inflation,
in each of which we explicitly use a shift symmetry as in supergravity- or superstring-based single-field inflation.
Both of the illustrative models contain a complex scalar together with only two free parameters. 
Sec.4 is devoted to a numerical calculation of the primordial curvature perturbation in these models,
where we will report their current status and prospects in light of future CMB-S4 experiments \cite{Abazajian:2016yjj}. 
Finally, we conclude in Sec.5.

\newpage
\section{Inflation In 2D Field Space}
\label{sec2}
In this section we briefly review two-field inflation with the two-dimensional (2D) field space composed of $\phi_{I}$, $I=1,2$. 
We will clarify the circumstances where a conservation of the primordial curvature (adiabatic) perturbation is violated by the presence of the isocurvature (entropy) perturbation.

\subsection{Backgrounds}
We begin with the two-field Lagrangian
\begin{eqnarray}{\label{Lag}}
S=\int d^{4}x \sqrt{-g}\left[\frac{M^{2}_{P}}{2}R-\sum_{I}g^{\mu\nu}\partial_{\mu}\phi_{I}\partial_{\nu}\phi_{J}-V(\phi_I)\right],
 \end{eqnarray}
where $g_{\mu\nu}$ is Friedmann-Robertson-Walker metric, $R$ is curvature, $M_P$ is reduced Planck mass,
$\phi_I$ denote the two independent scalar fields, and $V$ is inflation potential.
From Eq.(\ref{Lag}) the equations of the background fields are given as,
\begin{eqnarray}
3H^{2}&=&\frac{1}{2}\sum_{I}\dot{\phi}_{I}^{2}+V, \label{backe1}\\
\dot{H}&=&-\frac{1}{2}\sum_{I}\dot{\phi}_{I}^{2}, \label{backe2}\\
0&=&\ddot{\phi}_{I}+3\dot{\rho}\dot{\phi}_{I}+V_{I}, \label{backe3}
 \end{eqnarray}
where $H$ is Hubble parameter, ``dot" means derivative over time, and $V_{I}=\partial V/\partial \phi_{I}$.

To ensure an exponential expansion during inflation, 
 we introduce following slow-roll parameters as in the single-field case
\begin{eqnarray}
\label{ep}
\epsilon_{I}\equiv \frac{1}{2}\left(\frac{V_{I}}{V}\right)^{2}\approx  \frac{\dot{\phi}_{I}^{2}}{2H^{2}},
 \end{eqnarray}
where the equations of the background fields in Eqs.(\ref{backe1})-(\ref{backe3}) have been used.
Furthermore, 
in order to satisfy the slow-roll requirements $\mid \ddot{\phi}_{I}\mid<< 3H\mid\dot{\phi}_{I}\mid$,
we define the analogies of $\eta$ parameter
\begin{eqnarray}
\eta_{I}\equiv\frac{V_{II}}{V}\approx \epsilon-\frac{\ddot{\phi}_{I}}{H\dot{\phi}_{I}}-\delta^{IJ}\frac{V_{IJ}}{3H^{2}}\frac{\dot{\phi}_{J}}{\dot{\phi}_{I}}, \label{eta}
 \end{eqnarray}
where $V_{IJ}=\partial^{2}V/\partial \phi_{I}\partial \phi_{J}$ and $\epsilon=\sum_{I}\epsilon_{I}$.
The smallness of $\eta_{I}$ can be satisfied by imposing $\mid\dot{\phi}_{I}\mid\sim \mid\dot{\phi}_{J}\mid$ and $\mid V_{IJ}\mid <<H^{2}$ for $I\neq J$,
which are ``natural'' for the two-field inflation under consideration.
Because both scalar masses should be light compared to $H$,
and for $\mid\dot{\phi}_{I}/\dot{\phi}_{J}\mid$ either far larger or smaller than unity 
the effective number of the inflaton fields is actually reduced such as in quasi-single-field inflation \cite{Chen:2009zp, Noumi:2012vr}.

\subsection{Scalar Perturbations}
Let us now consider the scalar perturbations $\delta\phi_{I}$.
Along a classical trajectory in the 2D field space, 
one can transform the scalar perturbations as 
\begin{eqnarray}
\left(\begin{array}{c}
\delta\sigma\\
\delta s \\
\end{array}\right)=
\left(
\begin{array}{cccc}
\cos\theta& \sin\theta  \\
-\sin\theta &  \cos\theta  \\
\end{array}\right)
\left(\begin{array}{c}
\delta\phi_{1}\\
\delta\phi_{2} \\
\end{array}\right) \label{decom}
\end{eqnarray}
where $\cos\theta=\dot{\phi}_{1}/\sqrt{\dot{\phi}_{1}^{2}+\dot{\phi}_{2}^{2}}$ and $\sin\theta=\dot{\phi}_{2}/\sqrt{\dot{\phi}_{1}^{2}+\dot{\phi}_{2}^{2}}$. 
In terms of this transformation, 
the gauge-invariant curvature perturbation \cite{Bardeen:1983qw} reads
\begin{eqnarray}
\label{zeta1}
\zeta=-\frac{H}{\dot{\sigma}}\delta\sigma,
\end{eqnarray}
where $\dot{\sigma}=\cos\theta\dot{\phi}_{1}+\sin\theta\dot{\phi}_{2}$.
From Eq.(\ref{zeta1}) one finds that the new variables $\delta\sigma$ and $\delta s$ correspond to the curvature (adiabatic) and isocurvature (entropy) perturbation, respectively.

Following \cite{Garcia-Bellido:1995him,Finelli:2000ya,Gordon:2000hv}, the curvature perturbation in Eq.(\ref{zeta1}) on large scales evolves as
\begin{eqnarray}
\label{zeta2}
\dot{\zeta}=\frac{2H}{\dot{\sigma}}\dot{\theta}\delta s,
 \end{eqnarray}
Although the evolution of the curvature perturbation in Eq.(\ref{zeta2}) is a linear treatment,
it still reveals the condition for the non-conservation of the curvature perturbation.
Since the angle $\theta$ introduced in Eq.(\ref{decom}) characterizes the classical trajectory of inflation in the 2D field space,
$\zeta$ varies whenever it is not a straight line.
As a result, the curvature perturbation is altered due to the entropy perturbation either in inflationary\footnote{Under the slow roll conditions in Eq.(\ref{backe3}),
$\dot{\theta}\approx 0$ at the leading order of slow roll parameters, which seems to validate  the conservation of $\zeta$. However,  it doesn't hold at the next-leading order.} or post-inflationary\footnote{In reheating the role of the entropy perturbation is not only played by the scalar degrees of freedom of multifield inflation but also their decay products \cite{Taruya:1997iv,Jedamzik:1999um,Ivanov:1999hz,Finelli:2000ya}.} stage.

\section{Complex Field Inflation}
In this section we illustrate how to construct two kinds of complex field inflation in large-field contexts.

\subsection{Chaotic Inflation}
As mentioned above, complex chiral field is a building block of supersymmetry-, supergravity- or superstring-based inflation, see e.g, ref.\cite{1101.2488} for a review.
This topic is further divided into two different classes - the D-term \cite{Stewart:1994ts} and F-term \cite{1011.5945} inflation, where in the later case a realistic construction of inflation potential has to avoid the so-called ``$\eta$" problem. 

We restrict to a realization of complex chaotic inflation \cite{Linde:1983gd},
whose original two-field version is given as
\begin{eqnarray}
\label{chaotic1}
V(\phi_{I})=\sum_{I}\lambda_{I}\phi^{p}_{I},
 \end{eqnarray}
where $\lambda_I$ are two real parameters and the index $0< p \leq 4$. 
Instead of Eq.(\ref{chaotic1}), the potential for the complex chaotic inflation becomes
\begin{eqnarray}
\label{chaotic2}
V(\phi, \phi^{*})=\lambda\left(\phi^{p}+\phi^{* p}\right),
\end{eqnarray}
where there are only two model parameters $\lambda$ and $p$ as well.

To realize the inflation potential in Eq.(\ref{chaotic2}),
we make use of a shift symmetry as in refs.\cite{Kawasaki:2000yn,Takahashi:2010ky,Nakayama:2010kt} to avoid the well-known $\eta$ problem.
We consider a supergravity model with the following superpotential and Kahler potential
\begin{eqnarray}
\label{cL}
W&=&W_{0},\nonumber\\
K&=&\ln[-(z^{-n}+z^{* -n})]+\delta K(z, z^{*}),
\end{eqnarray}
where $z$ is the original complex scalar field,
$W_0$ is a constant term, $n$ is a real parameter with range $0<n<1$, 
the shift symmetry is identified as 
\begin{eqnarray}
\label{shift}
z^{-n}\rightarrow z^{-n}+ iR,
\end{eqnarray}
with $R$ a real parameter, 
and the small breaking term of the shift symmetry
\begin{eqnarray}
\label{dK}
\delta K(z, z^{*})=-\frac{c}{zz^{*}},
\end{eqnarray}
where $c$ a positive real parameter.
From Eq.(\ref{cL}) $c$ has to satisfy $c\mid z\mid^{-2}<n\ln(\mid z\mid)$.

In the parameter range $\mid z\mid >>1$ in unit of Planck mass, we have from Eq.(\ref{cL})
\begin{eqnarray}
\label{K}
K_{z}&=&\frac{\partial K}{\partial z}=\frac{cz^{*}}{(zz^{*})^{2}}-\frac{nz^{-n-1}}{z^{-n}+z^{*-n}},\nonumber\\
K_{z^{*}}&=&\frac{\partial K}{\partial z^{*}}=\frac{cz}{(zz^{*})^{2}}-\frac{nz^{*-n-1}}{z^{-n}+z^{*-n}}, \nonumber\\
K_{zz^{*}}&=&\frac{\partial^{2} K}{\partial z\partial z^{*}}=\frac{c}{(zz^{*})^{2}}-\frac{n^{2}(zz^{*})^{-n-1}}{(z^{-n}+z^{*-n})^{2}}.
\end{eqnarray}
So the $c$-terms in the above derivatives are the dominant contributions if one further imposes $c\mid z\mid^{-2}>n$.
Substituting Eq.(\ref{K}) into the supergravity Lagrangian \cite{1101.2488} yields
\begin{eqnarray}
\label{L}
\mathcal{L}\approx -\frac{c}{(zz^{*})^{2}}\sqrt{-g}g^{\mu\nu}\partial_{\mu}z\partial_{\nu}z^{*}-
\left(\frac{c}{zz^{*}}-3\right)\mid W_{0}\mid^{2}e^{\ln[-(z^{-n}+z^{* -n})]}.
\end{eqnarray}

Using a canonical normalization with $\phi=\sqrt{c}/z$
we finally obtain the complex chaotic potential 
\begin{eqnarray}
\label{V}
V\approx \lambda(\phi^{n}+\phi^{* n}),
\end{eqnarray}
with $\lambda=3c^{-n/2}\mid W_{0}\mid^{2}$ in the mildly fine-tuned parameter range $n<c\mid z\mid^{-2}<3$.
In this complex chaotic inflation, 
the super-Planck field range $\phi \sim \mathcal{O}(1)$ and 
the amplitude of the curvature perturbation can be adjusted by the range of $c$ and the magnitude of $W_{0}$ respectively.

\subsection{Exponential Inflation}
We now discuss a realization of complex exponential inflation.
Exponential inflation appears either in the context of supergravity \cite{Stewart:1994ts} or superstring \cite{Dvali:1998pa,Cicoli:2008gp}.
The two-field exponential potential takes a form
\begin{eqnarray}
\label{e1}
V(\phi_{I})=\sum_{I} V_{I}\left(1-e^{-q_{I}\phi}\right),
 \end{eqnarray}
where $V_I$ and $q_{I}$ are the model parameters.
In virtue of Eq.(\ref{e1}) the complex field version reads
\begin{eqnarray}
\label{e2}
V(\phi_{I})=V_{0}[\left(1-e^{-q\phi}\right)+\left(1-e^{-q\phi^{*}}\right)],
 \end{eqnarray}
where there are only two model parameters $V_0$ and $q$.

Here, we stick to a supergravity model with following Kahler potential 
\begin{eqnarray}
\label{eK}
K(\varphi,\varphi^{*})= \tanh^{-1}(\varphi)\tanh^{-1}(\varphi^{*}),
\end{eqnarray}
and Lagrangian 
\begin{eqnarray}
\label{eL}
\mathcal{L}=K_{\varphi\varphi^{*}}\sqrt{-g}g^{\mu\nu}\partial_{\mu}\varphi\partial_{\nu}\varphi^{*}-\lambda\left(\bar{\Psi}_{R}\varphi^{n}\Psi_{L}+\bar{\Psi}_{L}\varphi^{* n}\Psi_{R}\right),
\end{eqnarray}
where $\lambda$ is Yukawa coupling constant,
and $L$ and $R$ referring to the left- and  right-hand part of fermion $\Psi$ respectively.
Such Yukawa interaction can be either fundamental $(n=1)$ or arises from supergravity-induced effect $(n>1)$. 
For the fermion $\Psi$, we further assume that it is charged under a strong gauge dynamics with characteristic scale
\begin{eqnarray}
\label{cond}
\left<\bar{\Psi}\Psi\right>=\Lambda^{3},
\end{eqnarray}
where $\Lambda$ is assumed to be at least a few times of $M_P$.

To capture the low-energy effective theory 
we firstly determine the metric in terms of Eq.(\ref{eK}), which gives
\begin{eqnarray}
\label{metric}
K_{\varphi\varphi^{*}}=\frac{1}{1-\varphi^{2}}\frac{1}{1-\varphi^{*2}}.
\end{eqnarray}
Substituting Eq.(\ref{metric}) into Eq.(\ref{eL}) suggests a canonical normalization of the field 
\begin{eqnarray}
\label{rdef}
\varphi=\tanh(\phi/\sqrt{2}).
\end{eqnarray}
In the field range $M_{P}<\phi<\Lambda$,
one finds $\varphi\rightarrow 1$ in Eq.(\ref{rdef}) 
and the effective fermion mass is relatively light for $\lambda <<\Lambda$.
So, below $\Lambda$ scale the fermion condensate in Eq.(\ref{cond}) occurs and gives rise to 
 an effective potential 
\begin{eqnarray}
\label{eV}
V\approx V_{0}[(1-2ne^{-\sqrt{2}n\phi})+(1-2ne^{-\sqrt{2}n\phi^{*}})]
\end{eqnarray}
with $V_{0}=\lambda\Lambda^{3}$.
Note, in the field range $\varphi\rightarrow 1$,  $K^{-1}_{\varphi\varphi^{*}}\approx 0$ which suppresses the supergravity correction to Eq.(\ref{eV}).

Just like the previous examples, 
the overall coefficient $V_0$ is fixed by the amplitude of the primordial curvature perturbation,
whereas the other observables vary as the index $n$ changes.

\begin{table}
\begin{center}
\begin{tabular}{c}
\hline\hline
$\ln(10^{10}\mathcal{P}_{\zeta})=(3.044 \pm 0.014)$ (68 $\%$ CL)\\ 
 $n_{s}=0.965 \pm 0.004$ (68 $\%$ CL)   \\
$r< 0.10$ (95 $\%$ CL) \\
$f^{\rm{equil}}_{\rm{NL}}=-26 \pm 47$ (68 $\%$ CL) \\
\hline \hline
\end{tabular}
\caption{Constraints on observables of the primordial curvature perturbation from Planck 2018 data \cite{Planck:2018jri, Planck:2018vyg, Planck:2019kim}.}
\label{Planck}
\end{center}
\end{table}

\section{Numerical Analysis}
Having shown the viability of the model realizations of above two models of complex field inflation,
we proceed to discuss whether these models are compatible with current experimental data.
To calculate the observables of the primordial curvature perturbation in these models,
we use a public numerical code - transport approach \cite{Dias:2016rjq,Dias:2015rca}.
The transport approach, combined with PyTransport \cite{Mulryne:2016mzv}, 
is able to calculate observables of the primordial curvature perturbation such as 
power spectrum, tensor-to-scalar index and nonlinear parameter related to non-Gaussianty of bispectrum in multifield inflation whether the kinetic term for the complex scalar field is canonical or non-canonical.

The aforementioned observables of the primordial curvature perturbation,
which are strongly constrained by the latest Planck 2018 data as shown in Table.\ref{Planck},
are defined as following.
Firstly, the amplitude of the power spectrum is given by
\begin{eqnarray}
\label{PS}
\left<\zeta(\overrightarrow{k}_{1})\zeta(\overrightarrow{k}_{2})\right>=(2\pi)^{3}\delta^{(3)}(\overrightarrow{k}_{1}+\overrightarrow{k}_{2})P_{\zeta}(k)
\end{eqnarray}
with $\mathcal{P}_{\zeta}(k)=k^{3}P_{\zeta}(k)/(2\pi^{2})$, with the pivot scale $k_{*}=0.05$ Mpc$^{-1}$.
The explicit value of $\mathcal{P}_{\zeta}(k)$ fixes an overall magnitude of inflation potential.
Secondly, the spectrum index which describes the dependence of $\mathcal{P}_{\zeta}(k)$ on $k$ is defined as 
\begin{eqnarray}
\label{ns}
n_{s}-1=\frac{d\ln \mathcal{P}_{\zeta}(k) }{d\ln k}.
\end{eqnarray}
Thirdly, the tensor-to-scalar ratio is simply given by 
\begin{eqnarray}
\label{r}
r=\frac{P_{T} }{P_{\zeta}},
\end{eqnarray}
where $P_T$ is the tensor power spectrum. Finally, the nonlinear parameter $f_{\rm{nl}}$ which quantifies the non-Gaussianty of bispectrum with equilateral configuration reads 
\begin{eqnarray}
\label{fnl}
f_{\rm{nl}}=\frac{5}{18}\frac{B_{\zeta}(k)}{P^{2}_{\zeta}(k)},
\end{eqnarray}
where the bispectrum $B_{\zeta}(k)$ is given by
\begin{eqnarray}
\label{BS}
\left<\zeta(\overrightarrow{k}_{1})\zeta(\overrightarrow{k}_{2})\zeta(\overrightarrow{k}_{3})\right>=(2\pi)^{3}\delta^{(3)}(\overrightarrow{k}_{1}+\overrightarrow{k}_{2}+\overrightarrow{k}_{3})B_{\zeta}(k).
\end{eqnarray}

\subsection{Chaotic Inflation}
The parameter space of the complex chaotic inflation in Eq.(\ref{chaotic2}) is rather sensitive to the parameter $p$.
Because for certain value of $p$ the requirement of positivity of $V$ is so severe that viable field ranges are significantly suppressed, 
while for some $p$ this requirement is nearly absent.
To illustrate such feature in the complex chaotic inflation, 
in the following we consider two specific choices $p=\{0.25, 0.5\}$. 

\begin{figure}
\centering
\includegraphics[width=8cm,height=8cm]{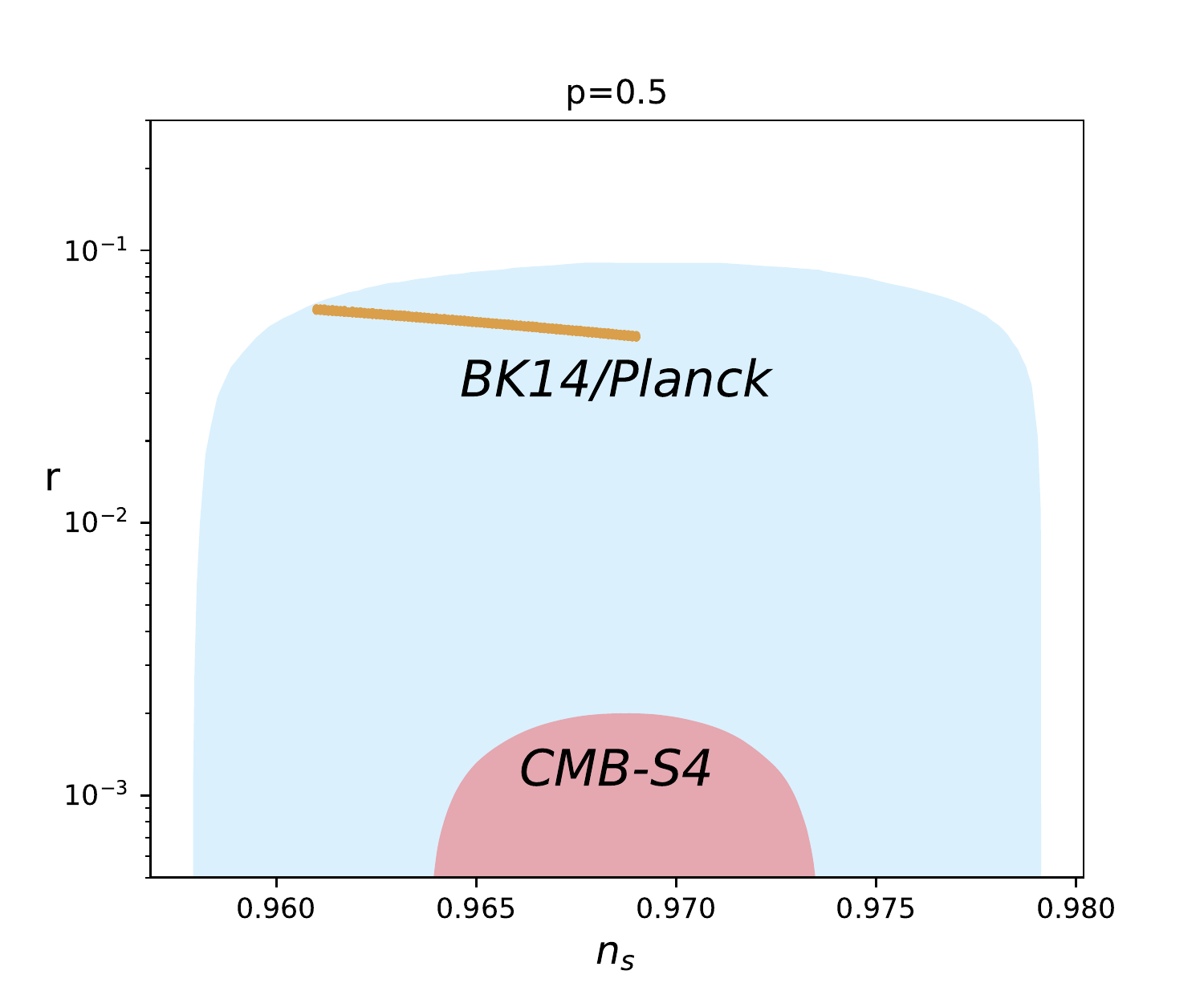}
\includegraphics[width=8cm,height=8cm]{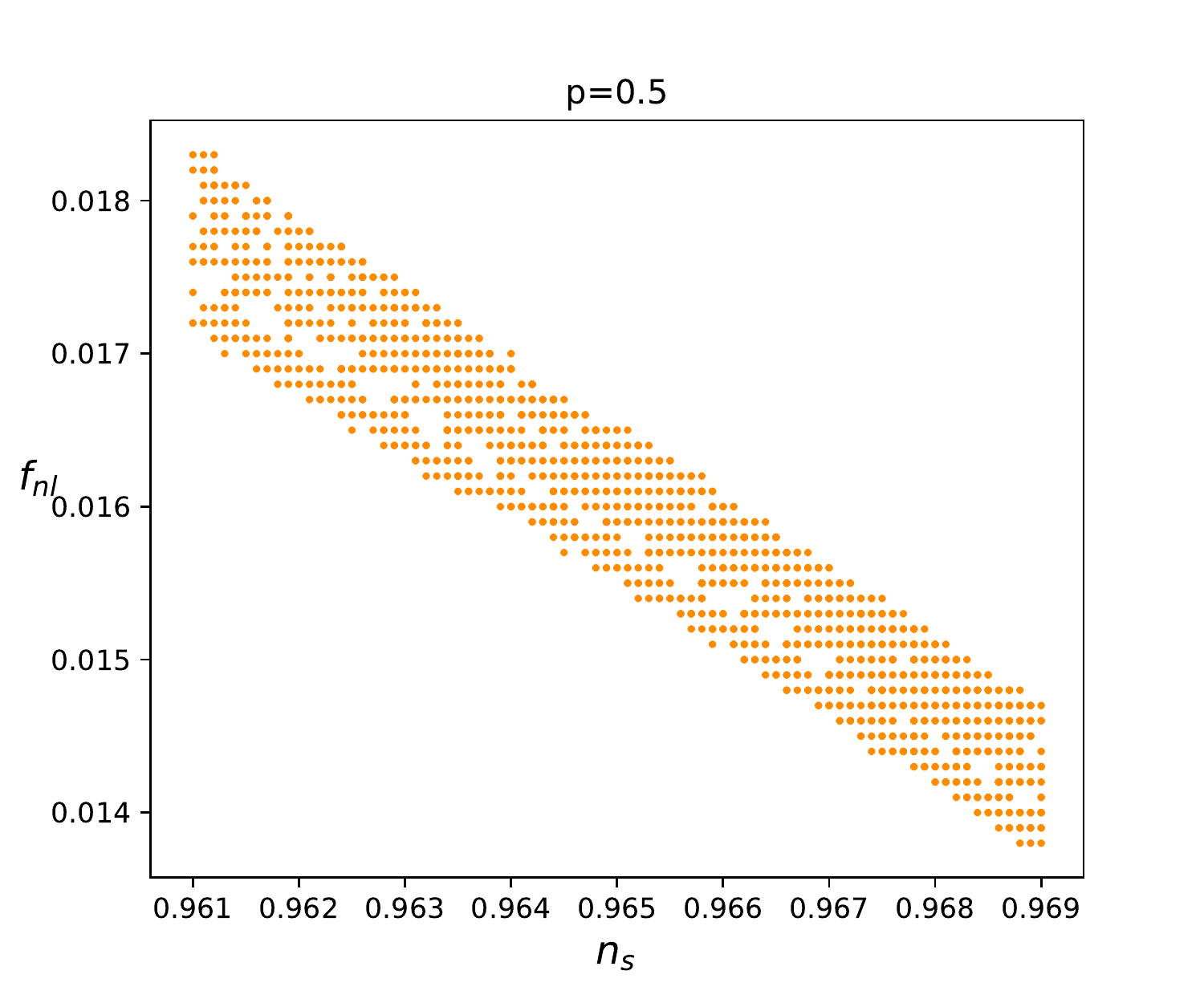}
\centering
\caption{Observables of the primordial curvature perturbation consistent with the Planck 2018 data in Table.\ref{Planck} 
in the complex chaotic inflation in the field ranges $\rm{Re}\phi\subset [3, 11]$ and $\rm{Im}\phi \subset [2, 11]$ for $p=0.5$, as compared with the constraint from BICEP2/Keck Array experiments plus Planck \cite{BICEP2:2015xme} (in blue) and the future CMB-S4 sensitivity \cite{Abazajian:2016yjj} (in red) at 95$\%$ CL.}
\label{co}
\end{figure}

Fig.\ref{co} shows the predictions of the primordial curvature perturbation consistent with current Planck 2018 data in Table.\ref{Planck} in the complex chaotic inflation in field ranges $\rm{Re}\phi\subset [3, 11]$ and $\rm{Im}\phi \subset [2, 11]$ for $p=0.5$.
This figure shows that the samples are also consistent with the constraint from BICEP2/Keck Array experiments plus Planck \cite{BICEP2:2015xme} (in blue) at 95$\%$ CL, 
and can be fully excluded by the future CMB-S4 sensitivity \cite{Abazajian:2016yjj} (in red) at 95$\%$ CL as they are distributed in a narrow region in the plane of $n_{s}-r$.
Compared to $p=0.5$, the case with $p=0.25$, which is not explicitly shown, 
has been confirmed to be excluded by the Planck 2018 data in Table.\ref{Planck}.

Combing the two cases illustrates that the predictions of the complex chaotic inflation obviously differ from those of the single-field chaotic inflation \cite{Planck:2018jri}, 
as both of the index choices in the later circumstance are still compatible with the Planck 2018 data in Table.\ref{Planck}.

\subsection{Expotential Inflation}
The parameter space of the complex exponential inflation in Eq.(\ref{eV}) is shaped by two facts.
Earlier we have seen a fact that the initial value of $\phi$ 
has to be at least a few times of Planck mass.
Another fact is that given an explicit value of the index $n$ the initial value of $\phi$ has to be upper bounded, as too large $\phi$ results in too small slow roll parameters for the scale $k_{*}$, 
which in turn yields $n_s$ too close to unity.
We expect that when $n$ increases the critical upper bound value of $\phi$ decreases, 
leading to a shrinking parameter space.
To manifest this point, we consider two specific cases with $n=\{0.5,1\}$.

\begin{figure}
\centering
\includegraphics[width=8cm,height=8cm]{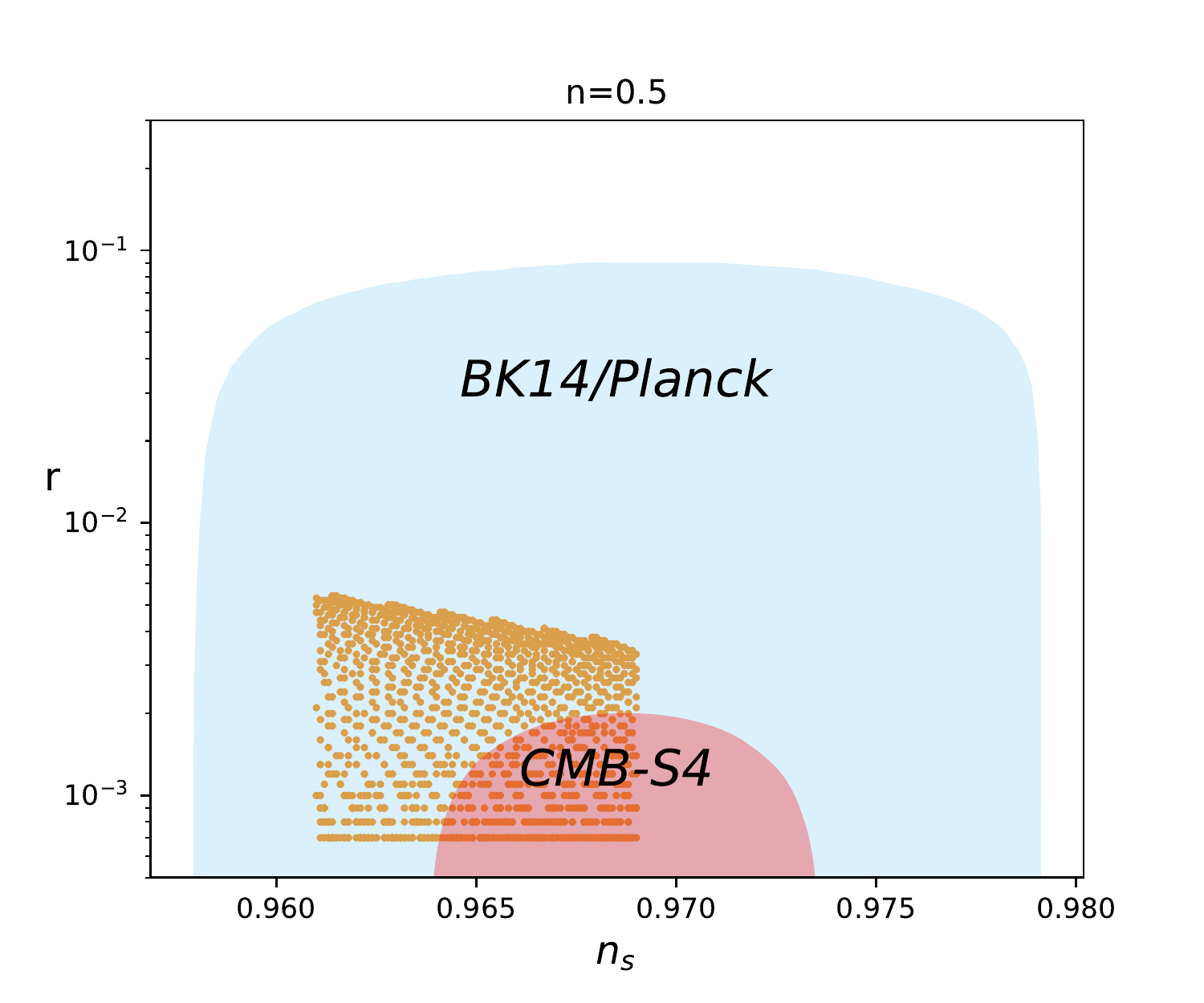}
\includegraphics[width=8cm,height=8cm]{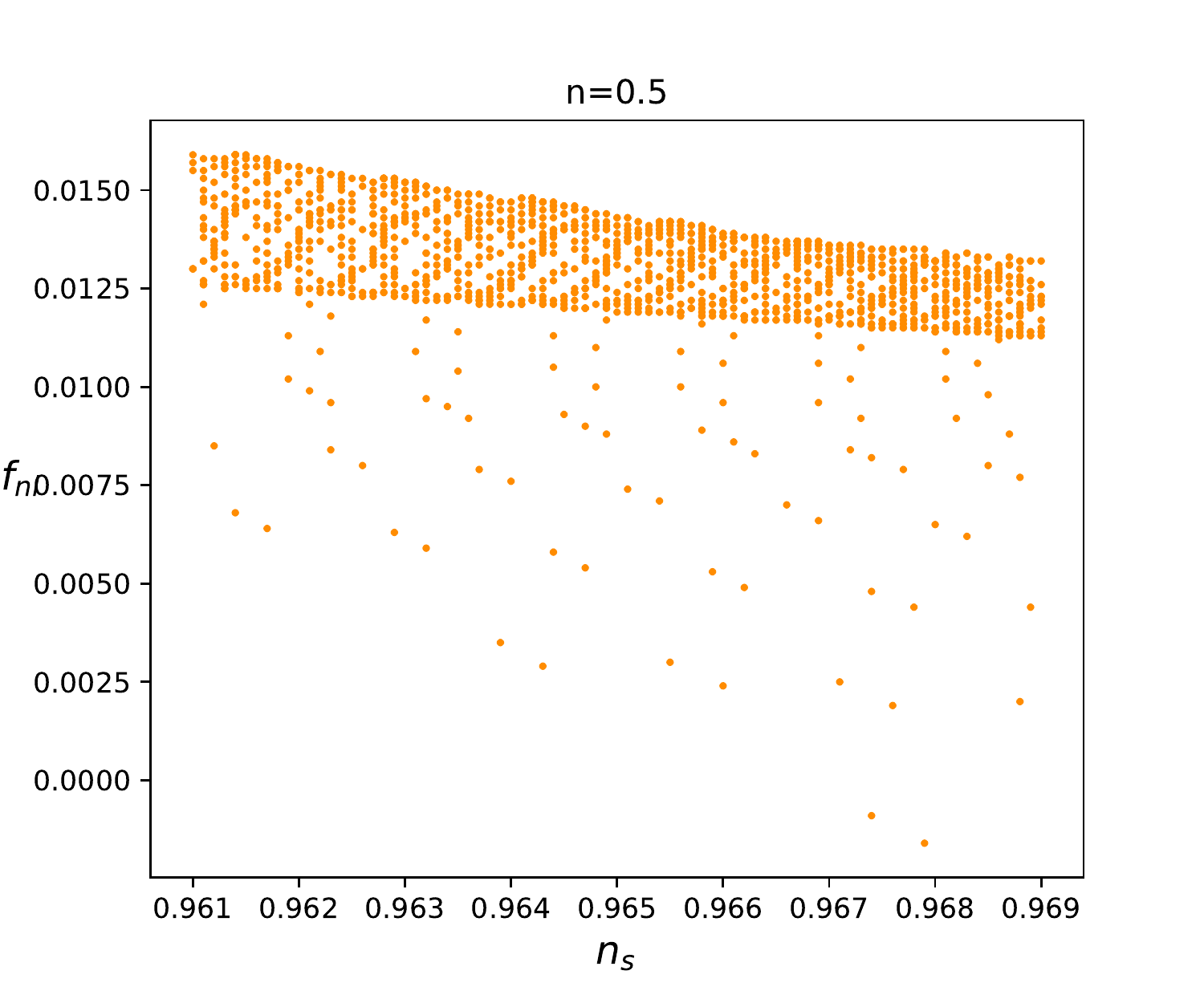}
\centering
\caption{Observables of the primordial curvature perturbation consistent with the Planck 2018 data in Table.\ref{Planck} 
in the complex exponential inflation in the field ranges $\rm{Re}\phi\subset [4, 8]$ and $\rm{Im}\phi \subset [4,12.9]$ for $n=0.5$.
The experimental limits are the same as in Fig.\ref{co}.}
\label{eo1}
\end{figure}

\begin{figure}
\centering
\includegraphics[width=8cm,height=8cm]{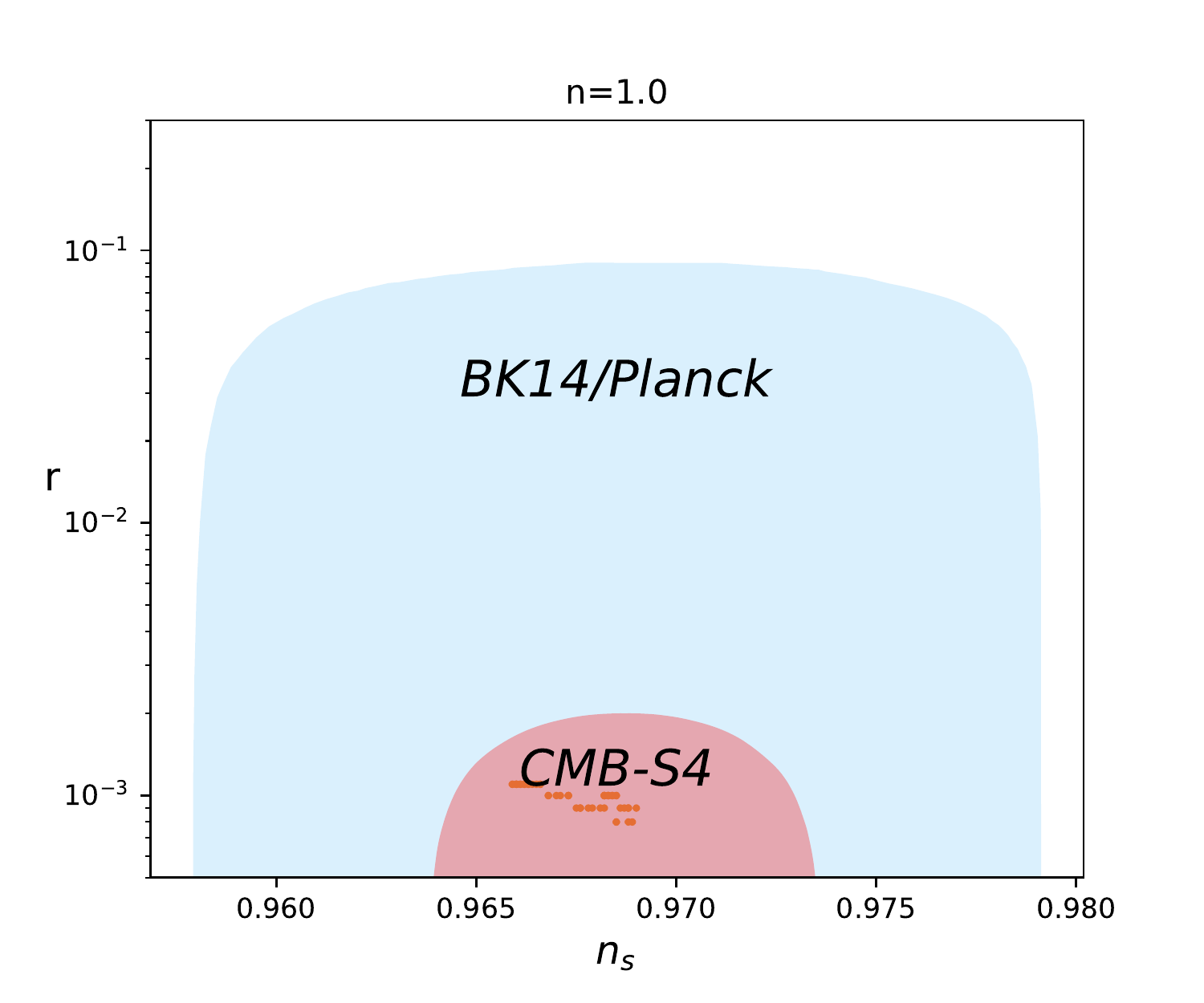}
\includegraphics[width=8cm,height=8cm]{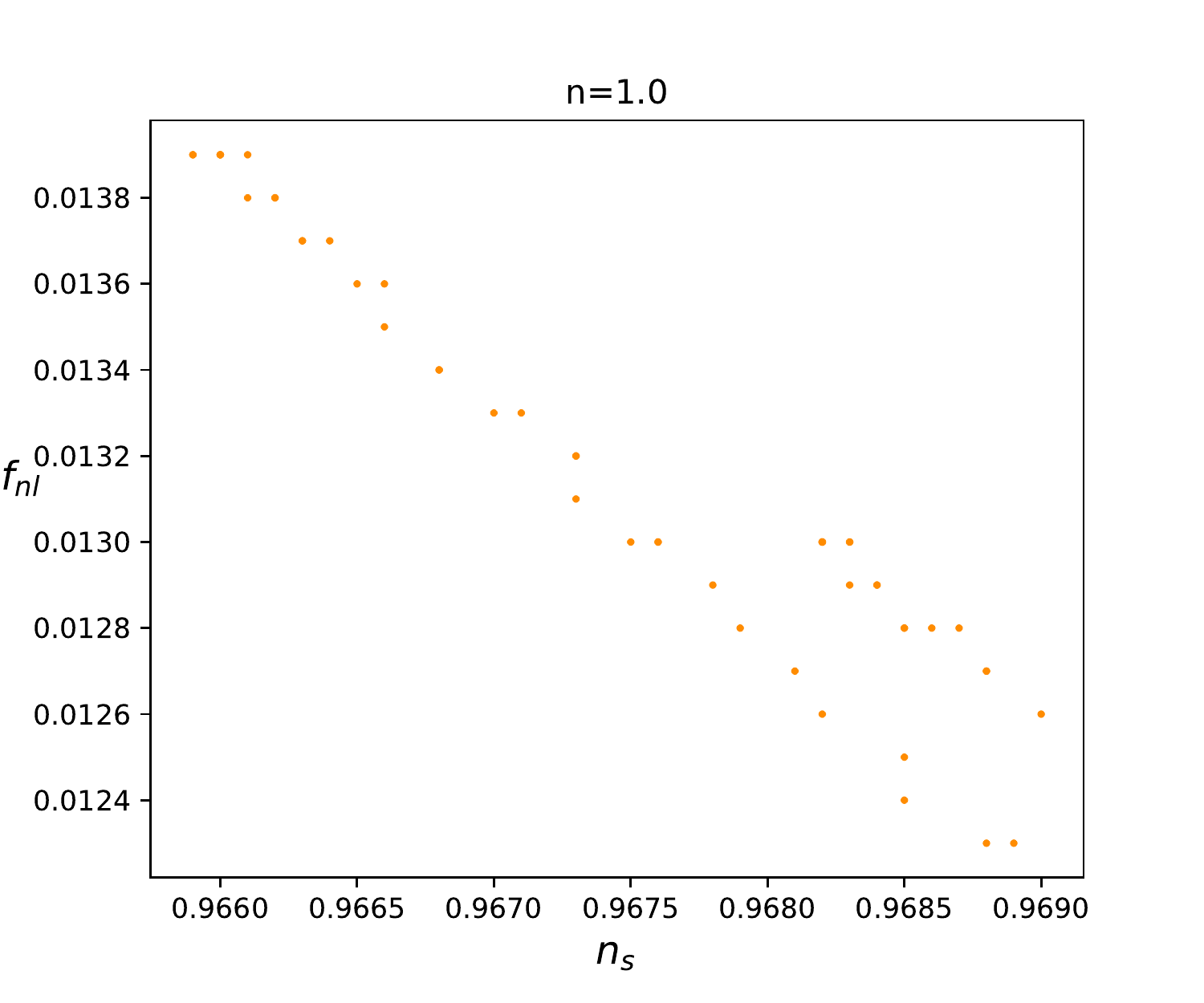}
\centering
\caption{Same as Fig.\ref{eo1} in the field ranges $\rm{Re}\phi\subset [4, 8]$ and $\rm{Im}\phi \subset [4,8.5]$ for $n=1$ instead.}
\label{eo2}
\end{figure}

Fig.\ref{eo1} and \ref{eo2} show the predictions of the primordial curvature perturbation for $n=0.5$ and $n=1$ respectively, 
where both of the field ranges are chosen to cover a complete period in the complex field space.
Explicitly,  
the number of samples is of order $\sim 100$ and $\sim 10$ for the case with $n=0.5$ and $n=1$ respectively.
The trend in the number of the samples implies that the upper bound value of $n$ is close to $1$.
As for observation, Fig.\ref{eo1} tells that 
the parameter space in the case with the smaller $n=0.5$ can be substantially excluded by the future CMB-S4 limit,
while Fig.\ref{eo2} shows that the parameter space with respect to the larger $n=1$  is beyond the reach of the same CMB-S4 limit.

In the light of the phenomenological study on the two models we have verified that $i)$
complex field inflation can indeed make robust predictions of the primordial curvature perturbation,
$ii)$ complex field inflation can be consistent with current Planck data,
and $iii)$ distinctive signal of complex field inflation, 
which differs from the original single-field inflation, 
can be tested by the future CMB measurements on $n_s$ and $r$, 
despite that the magnitude of $f_{\rm{nl}}$ is probably too small to be detected.

\section{Conclusion}
\label{sec5}
This work is devoted to advocate complex field inflation 
which makes more robust predictions on the primordial curvature perturbation than its original two-field version. 
In order to illustrate the viability of model realization of complex field inflation,
we have considered two representative examples, i.e, the complex chaotic and exponential inflation.
Similar to the supergravity- or superstring-based single field model, 
a shift symmetry has been used.
We then numerically calculate the predictions of the primordial curvature perturbation in these models based upon on the numerical code transport approach. 
It turns out that both two models are compatible with the latest Planck limits.
The distinctive signal in individual surviving parameter space,
which is different from the single-field prediction, 
can be substantially or fully probed by the future CMB experiments.

Along this study there are several directions which deserve further investigations.
In this work a model realization suffers from the fine-tuning issue,
which may be improved in circumstances due to an alternative shift symmetry instead of the adopted one.
Also, more examples of complex field inflation can be constructed especially in the small-field context which we did not consider here.
Lastly, there is an important technique needed to be developed. 
Here, we have implicitly assumed an  instantaneous reheating which follows the inflationary stage. 
Strictly speaking, the primordial curvature perturbation is inevitably changed by a reheating process which is highly likely to be nonlinear.
To our knowledge there is no public numerical tool to handle a general reheating process.

\section*{Acknowledgments}
The authors thank David Mulryne for correspondence about the numerical code transport approach. 
This research is supported in part by the National Natural Science Foundation of China under Grant No. 11775039.

\end{document}